\documentclass[a4paper,11pt]{article}
\pdfoutput=1
\usepackage{jcappub}
\usepackage{aas_macros}
\usepackage[T1]{fontenc}

\newcommand{\unit}[2]{#1\,\mathrm{#2}}
\newcommand{\unittwo}[2]{#1\,\,[\mathrm{#2}]}

\newcommand{\meq}[2][XXX]{\begin{equation}               
\label{eq:#1}
#2
\end{equation}}
\newcommand{\lr}[1]{\left({#1}\right)}
\newcommand{\lrsq}[1]{\left[{#1}\right]}				
\newcommand{\pard}[2]{\frac{\partial #1}{\partial #2}}
\newcommand{\dd}[0]{\ensuremath{\mathrm{d}}}
\newcommand{\Htwo}[0]{H${}_2$}%
\newcommand{\Hplus}[0]{H$\mathstrut^{+}$}
\newcommand{\Cplus}[0]{C$\mathstrut^{+}$}

\makeatletter
\g@addto@macro\bfseries{\boldmath}
\makeatother

\title{\boldmath Self-confinement of low-energy cosmic rays around supernova remnants}

\author[a]{Hanno Jacobs,}
\author[a]{Philipp Mertsch,}
\author[a]{ and Vo Hong Minh Phan}

\affiliation[a]{Institute for Theoretical Particle Physics and Cosmology (TTK),\\RWTH Aachen University, 52056 Aachen, Germany}

\emailAdd{jacobs@physik.rwth-aachen.de}
\emailAdd{pmertsch@physik.rwth-aachen.de}
\emailAdd{vhmphan@physik.rwth-aachen.de}

\abstract{Supernova remnants have long been considered as a promising candidate for sources of Galactic cosmic rays. However, modelling cosmic-ray transport around these sources is complicated by the fact that the overdensity of cosmic rays close to their acceleration site can lead to self-confinement, that is the generation of turbulence upon which these particles scatter. Such a highly non-linear problem can be addressed by numerically solving the coupled differential equations describing the evolution in space and time of the escaping particles and of the turbulent plasma waves. In this work, we focus essentially on the propagation of cosmic rays from supernova remnants in the warm ionized and warm neutral phases of the interstellar medium and propose an extended framework to take into account also the effect of energy loss relevant for cosmic rays of energy below $\unit{10}{GeV}$. Interestingly, the diffusion coefficient of low-energy cosmic rays could be suppressed by up to 2 orders of magnitude for several tens of kiloyears after the escape from the shock. The cosmic-ray spectrum outside the supernova remnant flattens below $\unit{1}{GeV}$ at a sufficiently late time reminiscient of the spectral behaviour observed by Voyager. We also find the grammage accumulated around the source to be non-negligible, with important implications for precision fitting of the cosmic-ray spectra.}

\notoc

\begin{document}

\hfill{TTK-21-56}

\maketitle
\flushbottom

\section{Introduction}
\label{sec:intro}

Supernova remnants (SNRs) have long been suspected~\cite{ZwickyBaade} as the most likely candidate for the bulk of Galactic cosmic rays (CRs).  
Recently, this has been supported by gamma-ray data~\cite{Fermi-LAT:2013iui,Jogler:2016lav} showing evidence for acceleration of hadronic CRs in a number of sources. 
Yet, the identification of any individual object as a source of the locally observed CRs is still elusive. 
Observationally, this identification is hampered by the fact that CR diffuse by scattering with turbulent magnetic fields, thus erasing almost all angular information. 
Theoretically, the acceleration of CR to the highest energies observed, their escape from the sources and their subsequent transport in the Galaxy need to be understood better (see e.g. \cite{grammage_Galactic_Gabici} for a review). 

In the last couple of years, it has become increasingly clear that the self-generation of turbulence is a process that is important not only for the acceleration, but also for the transport in the vicinity of sources and in the Galaxy at large: 
Close to their acceleration sites, the gradient formed by the overdensity of these particles introduces an additional term in the dispersion relation of plasma waves, leading to a growth of those waves travelling in the direction of the gradient~\cite{Kulsrud+2020}. In other words, CRs can generate the magnetic turbulence upon which they scatter, thus becoming self-confined around their sources (see \cite{Marcowith_2021} for a recent review). 
CR propagation, in this case, turns out to be a highly non-linear problem which is difficult to solve analytically. Cases in which the CR density is low enough to ignore the non-linearity are called test-particle cases.

What make the modelling difficult is the huge hierarchy of scales involved. 
Shock acceleration to the highest energies relies on the amplification of turbulent magnetic fields through streaming instabilities~\cite{Bell_2004,Amato_2009}, involving scales as small as the CR gyro radii of trans-thermal particles and is most reliably approached with computationally expensive particle-in-cell (PIC) simulation. 
The galactic transport, in contrast, is taking place on scales of kiloparsecs and is usually addressed by solving transport equations, either in analytical approximations or with numerical finite difference codes. 
The escape from the source and the transport of CR in their vicinity takes place on an intermediate scale where PIC simulations would be too expensive to run and the usual transport schemes are not reliable. 

To simplify the problem and focus on the escape and propagation of CRs, Malkov et. al.~\cite{Malkov2013} introduced the concept of a CR cloud. 
The CR cloud model considers the coupled dynamics of CRs and magnetic turbulence \emph{after} acceleration of CRs by that source. 
Instead of emulating the detailed, but model-dependent results of shock acceleration in terms of spatial and spectral distributions, the CRs are assumed to be homogeneously distributed in a limited spatial region, the CR cloud, and follow a power law spectrum. 
Malkov et. al. analytically derived a self-similar solution for the system of transport equations under the assumption of negligible damping of turbulence. 
Subsequently, Nava et al. \cite{Nava_2016,Nava_2019} solved the equations numerically including the effect of damping, also providing an estimate for the escape time.
Most recently Brahimi et al. \cite{Brahimi2020} investigated the propagation into the inhomogeneous interstellar medium and Recchia et al. \cite{Recchia_2021} refined the coupling of ions and neutrals for particles down to $\unit{10}{GeV}$.

Most of the previous studies have focused on the energy range above $\unit{10}{GeV}$ while the peak of the CR energy density is known to be at lower energies than that. 
Thus, in principle, self-confinement could be be stronger at smaller energies than considered previously. 
While historically, the modelling of CRs at energies lower than a few GeV was limited by the scarcity of data, recently the Voyager probes have left the solar system and can now measure the fluxes of low-energy CRs without being affected by the solar wind~\cite{Voyager_out,Voyager_data}. 
It was soon realised that a simple extrapolation of AMS-02 data from higher energies fails and a spectral break in the source spectrum is needed to explain the Voyager data around $\unit{100}{MeV}$ in the standard model of CR propagation \cite{Boschini_2018a,Boschini_2018b,Orlando_2018,source_break,Bisschoff_2019}. 
In addition, the stochastic nature of CR sources was recently also shown to affect the spectra of CRs due to the small loss-lengths of low-energy CRs \cite{Phan_2021}. 

We stress that low-energy CRs might be even more important for the dynamics of the ISM again as the peak of CR energy density is at about $\unit{1}{GeV}$. Since these particles are energetic and abundant enough to penetrate and ionise the interior of molecular clouds, they determine the ionisation rate which ultimately controls not only the chemistry but also the coupling of gas and magnetic fields in star forming regions \cite{Padovani_2009, Morlino_2015,Ivlev_2015,Phan_2018}. 
It is not clear yet whether or not SNRs are also principal sources of low-energy CRs. Nevertheless, the enhanced ionisation rates inferred from sub-millimeter and infrared observations of molecular clouds interacting with SNRs suggest the presence of low-energy CRs originating from these objects (see e.g. analyses of clouds around SNR IC443, W51, and W28 \cite{Indriolo_2010,Ceccarelli_2011,Vaupre_2014,Phan_2020}). 

Answering the above questions thus requires a better modelling of self-confinement of CRs below $\unit{10}{GeV}$ around SNRs. 
However, low-energy CRs undergo significant energy losses which the previously used setups could not cover \cite{Nava_2016,Nava_2019,Recchia_2021} due to the choice of numerical methods used.  
Here, we will propose an extended framework which will be able to take into account the energy losses which allows us to model the self-confinement of CRs below $\unit{10}{GeV}$. 
Additionally, we will improve on the spectral transport of magnetic turbulence.

The remainder of the paper is structured as follows: 
In section \ref{sec:model} we describe our model. We explain the transport equations,  both for CRs and magnetic turbulence. 
We also review the physical processes at work, extending on the work of \cite{Recchia_2021} by introducing energy losses and the turbulent cascade.
Our results will be shown and discussed in section \ref{sec:results}. 
Finally, in section \ref{sec:conc} we conclude and give a short outlook.

\section{Model}
\label{sec:model}

We model the transport of CRs in the source vicinity in the one-dimensional, so-called flux-tube approximation. 
This is based on the fact that on length scales much smaller than the coherence length $L_c\leq \unit{100}{pc}$ \cite{LC_Beck}, magnetic field lines are nearly parallel. 
Since the parallel component of the diffusion tensor is larger than the perpendicular one, particles diffuse predominantly in one dimension along the magnetic field lines. 
On scales much larger than $L_c$, field line diffusion makes the orientation of the large-scale magnetic field more uniform and diffusion is isotropic. 
The higher dimensionality on large scales also leads to a faster suppression of fluxes through diffusion. 
While a proper modelling of this transition is beyond the scope of this paper, in the flux tube approximation, this can be emulated by a free escape boundary at either end of the flux tube. 
In the following we choose $L_c=\unit{100}{pc}$.

\subsection{Transport equation}

The propagation of CR protons in a 1D flux tube is described by the usual transport equation~\cite{Jokipii_1966,Blandford_1987}:
\begin{equation}
    \frac{\partial f}{\partial t} +v_{\rm A}\frac{\partial f}{\partial z}-\frac{\partial}{\partial z}\left[D(z,p,t)\frac{\partial f}{\partial z}\right]-\frac{dv_{\rm A}}{dz}\frac{p}{3}\frac{\partial f}{\partial p} + \frac{1}{p^2}\frac{\partial}{\partial p}\left[p^2\lr{\frac{\dd p}{\dd t}} f\right]=Q_p(z,p,t) \, , \label{eq:tpep}
\end{equation}
where $f = f(z, p, t)$ is the phase space density of protons. The second term on the left hand side of eq.~\eqref{eq:tpep} describes advection at the (local) Alfvén speed $v_{\rm A}$, the third one diffusion with the parallel diffusion coefficient $D(z,p,t)$. Momentum losses by adiabatic expansion are accounted for by the fourth term and other energy loss processes, e.g.\ due to ionisation or pion production, by the last term. The term $Q_p(z,p,t)$ on the right hand side describes sources of CRs.

The diffusion of CRs is mediated by scattering processes on the turbulent magnetic field with turbulence spectral power $W = W(z, k, t)$. We consider Alfv\'en waves of wavelength $(2 \pi) / k$ as the main scattering centres for CRs with Larmor radius $r_ {\rm L}(p) = 1/k$, that is the interactions are resonant. In this case the diffusion coefficient is given by~\cite{Bell_1978}:
\begin{equation}
\label{eq:dc}
    D(p, z, t)=\frac{D_{\rm B}(p) 4/\pi}{kW(k,z,t)},
\end{equation}
where $D_{\rm B}(p)=r_ {\rm L}(p)c\beta/3$ is the Bohm diffusion coefficient. This couples eq.~\eqref{eq:tpep} to the transport equation for the turbulence spectral power~\cite{TEW}:
\begin{equation}
\label{eq:Weq}
    \frac{\partial W}{\partial t}+ \frac{\partial }{\partial z}(v_{\rm A}W) = \lr{\Gamma_{\rm CR}(f)-\Gamma_{D}(W)}W+\Gamma_{D}(W_{\rm BG})W_{\rm BG}+\pard{}{k}\left[D_{kk}(W)\pard{W}{k}\right] \, .
\end{equation}
Here, $\Gamma_{CR}(f)$ describes the growth of turbulence due to the resonant streaming instability, see section~\ref{subsec:ressi}, and $\Gamma_D$ are damping terms as explained in section \ref{subsec:damping}. 
The turbulent cascade is described by a diffusion in wavenumer $k$, in which eddies of similar size interact: 
A vortex will disturb the flow velocity and cause new vortices at a rate determined by the wavenumber diffusion coefficient $D_{kk}(W)$. 
Oftentimes, this is approximated with a damping term~\cite{Evoli_2018,Mukhopadhyay_2021}, as transfer of turbulent energy density takes place mainly towards smaller scales.
In Kolmogorov turbulence phenomenology the diffusion coefficient in $k$ is given by~\cite{DCk_Miller}:
\meq[DCkKOL]{D_{kk}(W)=c_kv_{\rm A}k^{7/2}\sqrt{W},}
and for Kraichnan turbulence by:
\meq[DCkKRA]{D_{kk}(W)=c_kv_{\rm A}k^4W,}
where $c_k=0.052$~\cite{Evoli_2018}.

In the absence of a CR gradient, the turbulence level has to retain the average ISM value. 
Since the continuous injection at large scales and dissipation at small scales is neglected in this work, this requires the introduction of the background compensating term $\Gamma_D(W_{BG})W_{BG}$.

\subsection{Energy losses}

The last term on the left hand side of eq. \eqref{eq:tpep} accounts for energy losses. 
It can be neglected for nuclei at higher energies as done in previous work, but will become important at lower energies as the loss time scales become comparable to or shorter than the other time scales involved. 
Here, we collect the energy loss rates we use. 

Protons interact with the ionised thermal background plasma via Coulomb interactions. 
A good approximation of this loss rate is given by eq. 4.22 of Ref.~\cite{Eloss_c_i} :
\meq[Eloss-c]{-\lr{\frac{\dd p}{\dd t}}_{\rm Coulomb}\approx 3.1 \times 10^{-7}\lr{\frac{n_e}{\mathrm{cm}^{-3}}}\frac{\beta}{x_m^3+\beta^3}\mathrm{\frac{eV/c}{s}},}
with $x_m=0.0286\lr{T/\unit{2\times 10^6}{K}}$, the electron number density $n_e$ and the particle speed $\beta=v/c$. 

Protons also lose energy by ionising neutral particles.
The momentum loss rate is given by eq.~4.32 of Ref.~\cite{Eloss_c_i}:
\meq[Eloss-i]{-\lr{\frac{\dd p}{\dd t}}_{\rm ion}\approx 1.82 \times  10^{-7}\lr{\frac{n_{\rm H}}{\mathrm{cm^{ -3}}}}\lrsq{1+0.185 \, \mathrm{ln}(\beta)\theta(\beta-\beta_0)}\frac{2\beta}{\beta_0^3+2\beta^3}\mathrm{\frac{eV/c}{s}},}
where $\beta_0=0.01$, $\theta$ is the Heaviside function and $n_H$ the number density of hydrogen. 
Here we have fixed a typo in eq.~4.32 of Ref.~\cite{Eloss_c_i}, where the factor in front of the logarithm is one order of magnitude too small. 
(Compare with their eq.~4.31.) This equation is valid for energies above $\unit{100}{keV}$.

At energies higher than $\unit{1}{GeV}$ CR protons lose energy when they create pions in the interaction with protons of the ISM.
The energy loss rate of this proton-proton interaction is given by eq. 34 of Ref. \cite{Eloss_pp}:
\begin{equation}
    \label{Eloss-p}
    -\lr{\frac{\dd p}{\dd t}}_{\rm pp}\approx 3.85 \times 10^{-7}\lr{\frac{n_{\rm H}}{\mathrm{cm^{ -3}}}}\lr{\frac{E}{{\rm GeV}}}^{1.28}\lr{\frac{E}{{\rm GeV}}+200}^{-0.2}\beta^{-1}\mathrm{\frac{eV/c}{s}}.
\end{equation}
All three processes combined give the total energy loss rate.

\subsection{Resonant streaming instability}
\label{subsec:ressi}

If the drift velocity of CRs 
\begin{equation}
    \label{eq:vd}
    v_{\text{d}}=-\frac{D}{f}\frac{\partial f}{\partial z}
\end{equation}
is larger than the local Alfvén speed $v_{\text{A}}$, CRs transfer energy to the waves.
In kinetic theory this can be derived considering the contribution of CR to the dispersion relation of plasma waves ~\cite{Krall_Trivelpiece}.
This additional term leads to an imaginary part in the wave frequency and, depending on the wave helicity, wave direction and CR gradient direction, this leads to damping or growth of the wave amplitude. 
In the latter case, this is called the CR streaming instability (see e.g. \cite{ressi_protons_Achterberg}).

The resulting growth rate of $W$ can be shown to be~\cite{1971ApJ...170..265S}
\begin{equation}
\label{eq:ressi}
    \Gamma_{\rm CR}(f)=-\frac{4 \pi}{3}\frac{c \left\vert v_{\rm A} \right\vert}{k W(k) U_0}\beta(p) p^{4}\frac{\partial f}{\partial z},
\end{equation}
where $U_0=B_0^2/8\pi$ is the energy density of the background magnetic field. 
Note that due to the assumption of resonant interactions, there is a one-to-one relation between wavenumber $k$ and particle Larmor radius $r_ {\rm L}(p)$, i.e.\ $k=1/r_ {\rm L}(p)$. 
In this work we only consider the scattering upon forward propagating waves and we ignore the presence of backwards propagation waves.

\subsection{The interstellar medium}

The expansion of the SNR as well as the Alfvén speed and the damping of turbulence depend on the ambient environment. 
Commonly, five different phases of the interstellar medium (ISM) are distinguished. 
These range from the hot ionised medium (HIM), found in stellar bubbles, over the warm ionised medium (WIM) and warm neutral medium (WNM) to the denser phases in or around molecular clouds, the cold neutral medium (CNM) and diffuse molecular medium (DiM)~\cite{ISM_2001,ISM_2019}. 
While the colder and denser phases contribute most of the mass in the ISM, those also have the smallest filling factor. 
Supernovae, instead, most often explode in the media with larger filling factors, as is also hinted at by their composition \cite{ISM_CR_COMP}. 
For this reason, the CNM and DiM are not considered in this study.
All relevant properties for this work are shown in table~\ref{tab:ism}.

The ionisation fraction $f_i$ and the helium fraction $\chi$ are defined as
\begin{equation}
    \label{eq:f_chi}
    f=\frac{n_i}{n_{\rm H}+n_i}, \qquad \chi=\frac{n_{\rm He}}{n_{\rm H}+n_i}, 
\end{equation}
where $n_{\rm H}$ is the number density of hydrogen atoms, $n_i$ the ion number density, and $n_{\rm He}$ the number density of helium atoms.

\begin{table}[t]
    \centering
    \begin{tabular}{| c|c|c|c|c|c|c|c |}
    \hline
    ISM & $\unittwo{T}{K}$  & $\unittwo{n}{cm^{-3}}$ &   $f_V$   & $f_i$   & $\chi$ &   neutrals & ions \\
    \hline\hline
				HIM		& $10^6$		& $10^{-2}$	& 0.5			& 1		& $0$	& - 			& \Hplus{} \\%
				\hline
				WIM		& $8000$		& 0.35 		& 0.25		& 0.6-0.9		& 0.1 & H, He		& \Hplus{} \\%
				\hline
				WNM	& $8000$		& 0.35		& 0.25		& $10^{-2}$	& 0.1 & H, He		& \Hplus{} \\%
				\hline
				CNM		& $80$		& 35			& $\sim 0$	& $10^{-3}$	& 0.1 & H, He		& \Cplus{} \\%
				\hline
				DiM		& $50$		& 300		& $\sim 0$	& $10^{-4}$	& 0.1 & \Htwo{}, He	&\Cplus{}\\
				\hline
    \end{tabular}
    \caption{Properties of the different phases of the ISM~\cite{ISM_2001,ISM_2019,Recchia_2021}. $T$ is the temperature of the medium, $n$ the total number density, $f_V$ the Galaxy filling factor, $f_i$ the ionisation fraction and $\chi$ the He fraction. The magnetic field in all phases is $B_0=\unit{5}{\mu G}$.}
    \label{tab:ism}
\end{table}

\subsection{Damping of Alfvén waves}
\label{subsec:damping}

Depending on the phase of ISM considered, different damping processes can dominate. In the phases considered here the ion neutral, Farmer Goldreich and non-linear Landau damping are most important.

\subsubsection{Ion-neutral damping}

In partially ionised plasmas the ions that form the wave encounter collisions with neutral particles. 
Due to these interactions the wave will be dampened. 
The damping rate depends on how often collisions occur. 
If the collision frequency is much larger than the wave frequency, the neutrals are well coupled. 
The neutral mass density must then be taken into account in the Alfvén speed which makes it smaller than in the case with ions only. 
On the other hand, if the collision frequency is much smaller than the wave frequency the neutrals will not co-oscillate, but rather dampen the wave. 
In both cases the ion neutral damping dominates under most circumstances in plasmas with neutral fractions above $\unit{10}{\%}$~\cite{Recchia_2021}.  

If the ion-to-neutral mass density
\begin{equation}
    \epsilon = \frac{m_in_i}{m_nn_n}
\end{equation}
is small, $\epsilon \ll 1$, the damping rate is approximately given by
\begin{equation}
    \Gamma_D^{\rm in}\approx\frac{\omega_k^2\nu_{\rm in}}{2\left[\omega_k^2+(1+\epsilon)^2\nu_{\rm in}^2\right]}.
\end{equation}
The ratio of the ion neutral collision frequency
\begin{equation}
    \nu_{\rm in}=\frac{m_n}{m_i+m_n}\left< \sigma_{\rm mt} v\right>_{\rm in}n_n
\label{eq:vin}
\end{equation}
and the wave vector in units of the Alfvén speed $\omega_k=kv_{\rm A,i}$ determines the coupling of ions and neutrals. The cross sections
This sets both the Alfvén speed and the damping rate. 
In the following, we consider both coupling regimes in turn. 

\paragraph{Efficient coupling $\omega_k \ll \nu_{in}$:}
For high-energy CRs the resonant wave number is short compared to $\nu_{in}/v_{\rm A}$ and the neutrals are well coupled to the wave. 
Hence, the Alfvén speed depends on both ions and neutrals, 
\meq[vAn]{v_{\rm A,n}=\frac{B}{\sqrt{4\pi\mu m_p n}},}
where $\mu=\sum_{s}m_s n_s/(m_p n)$ for all species $s$ of the medium. 
In this limit, the damping rate is $\Gamma_d^{\rm in}\propto E^{-2}$.

\paragraph{Weak coupling $\omega_k\gg\nu_{in}$:}
Low-energy CRs are resonant with waves of high wave number. 
In this case the neutrals are unable to follow the movement of the ions and the Alfvén speed depends only on the ions:
\meq[vAi]{v_{\rm A,i}=\frac{B}{\sqrt{4\pi\mu m_p n_i}},}
where $n_i$ is the number density of ions and $\Gamma_d^{\rm in}\propto \mathrm{const}$.
This will increase the Alfvén speed and thereby the streaming instability at low energies, compared to higher ones.

\subsubsection{Turbulent damping}

Magnetic turbulence in the Galaxy is thought to be injected on large scales by SNRs, stellar winds or similar processes (see eg.~\cite{BGturb}) with a turbulent velocity $v_{\rm turb}$ at a scale $L_{\rm inj}$. 
The turbulent cascade is highly anisotropic elongating perturbations along the magnetic field lines. A wave of initially $\lambda_{\perp}$ and $\lambda_{\parallel}$ will suffer an order unity shear after travelling a distance where the field lines spread by $\lambda_{\perp}$. When waves with $\lambda_{\perp}$ interact with oppositely directed Alfvén wave packets of the cascade, one wave period in the cascade corresponds to many periods of the wave and the Alfvén wave will experience the wave of the turbulent cascade as background. This leads to the cascade of the wave until the dissipation scale, and ultimately to the dissipation of turbulence.
The resulting damping rate of turbulence is called Farmer-Goldreich damping~\cite{FarmerGoldreich,YanLazarian} and given by:
\begin{equation}
    \Gamma_D^{\rm FG}=\lr{\frac{v_{\rm turb}^3}{L_{\rm inj}r_ {\rm L}v_{\rm A}}}^{1/2}.
\end{equation}

On the largest scales, given by the size of the injection regions and assumed to be $L_{\rm inj}=\unit{50}{pc}$, the waves are resonant with high-energy particles. 
Therefore, the injection velocity $v_{\rm turb}$ is given by $v_{ A,n}$, since neutrals are well coupled.

If the fration of neutrals is appreciable, ion neutral damping limits the cascade of the external turbulence. 
The Farmer Goldreich damping in this case has a lower limit in wave length  $l_{\rm min}=1/k_{\rm min}$, given by
\begin{equation}
    \label{eq:FGlimit}
    \frac{1}{l_{\rm min}}=L_{\rm inj}^{1/2}\lr{\frac{2\epsilon\nu_{\rm in}}{v_{ \rm A,n}}}^{3/2}\sqrt{1+\frac{v_{\rm A,n}}{2\epsilon\nu_{\rm in}L_{\rm inj}}.}
\end{equation}

\subsubsection{Non-linear Landau damping}

In regions of increased turbulence Alfvén waves are likely to interfere and form a beat wave. 
The modulation frequencies of this superposition are given by the sum and the difference of the individual wave frequencies. 
If both waves have similar frequencies, the group velocity of the beat $v_{\rm gr}$ can be close to the velocity of thermal background ions $v_{\rm pl}$. 
Waves will transfer energy to the plasma if $v_{\rm gr}>v_{\rm pl}$ and vice versa. 
Since there exist more lower energy than higher energy background particles, the net effect is a damping. 
For a single wave, this is called Landau damping and in the case of a beat wave called non-linear Landau damping. 
The damping rate is given by~\cite{nlld}
\begin{equation}
    \label{eq:nlld}
    \Gamma_D^{\rm NLLD}=\sqrt{\frac{\pi}{2}\lr{\frac{k_BT}{\mu m_p}}}\frac{W(k)}{r_ {\rm L}^2(p)},
\end{equation}
where $k_B$ is the Boltzmann constant, $T$ the temperature of the thermal background and $\mu m_p$ the effective mass.

\subsection{Initial conditions}

In the model of a CR cloud particles are initially confined within the SNR. 
In order to motivate this idea and explain the initial release time and radius it is necessary to understand the underlying physical phenomena.
When a massive star explodes at the end of its lifetime it ejects large amounts of mass into the ISM. 
In the initial phase of the expansion, when the ejected mass $M_{\rm ej}=1.4M_{\odot}$ is larger than the mass of the interstellar material swept up by the shock, the expansion is free~\cite{SNRfree}.
Afterwards the SNR expands adiabatically in the Sedov-Taylor phase and the radius can be described by~\cite{R_ST}
\begin{equation}
\label{eq:RST}
    R_{\rm SNR}(t)=0.5\lr{\frac{E_{51}}{n}}^{1/5}\lr{1-\frac{0.009M_{\rm ej,\odot}^{5/6}}{E_{51}n^{1/3}t_{\rm kyr}}}^{2/5}t_{\rm kyr}^{2/5}\,\mathrm{pc},
\end{equation}
where $E_{51}$ is the supernova energy in $\unit{10^{51}}{erg}$, $n$ the ISM number density in $\unit{1}{cm^{-3}}$, $M_{\rm ej,\odot}$ the ejecta mass in units of solar masses and $t_{\rm kyr}$ the supernova age in $kyr$. 
In the outer shells cooling processes become important at~\cite{t_rad},
\begin{equation}
\label{eq:trad}
t_{\rm rad}\approx 1.4\times E_{51}^{3/14}n^{-4/7}\,\mathrm{yr} \, ,
\end{equation}
and the hotter inner part compresses the material near the edge. 
This marks the beginning of the radiative phase and slows down the expansion of the SNR.

Since the expansion into the ISM happens at supersonic speeds, a shock forms. 
At the shock particles are accelerated by non-linear diffusive shock acceleration to a power law in momentum, which at low energies is softer than $p^{-4}$ and turbulence is created. 
The exact values and mechanisms lie beyond the scope of this work. 
Models can be found in~\cite{Malkov_Drury_2001} or more recently \cite{nldsa_diesing_2021}. In the following the spectral index is chosen to be $\alpha=4.2$. 

In the radiative phase the shock speed reduces rapidly, ending efficient particle acceleration, and the increased density dampens the turbulence in this region. 
While higher energetic particles can escape earlier from the SNR \cite{Nava_2016}, lower energetic particles with a smaller Larmor radius can only escape at this time \cite{Recchia_2021}. 
This containment of low-energy particles and their simultaneous release during the radiative phases can be modelled by a step-like initial condition,
\begin{equation}
    f(z, p)=\begin{cases}
        f_0(p)\qquad&\mathrm{ if }\,z\leq R_{\rm SNR}(t_{\rm rad}),\\
        0 \qquad\ &\mathrm{ if }\,z>R_{\rm SNR}(t_{\rm rad}),
        \end{cases}
\end{equation}
where we take the momentum dependence to be power-law and the overall normalisation is set by the requirement that the total CR energy initially contained in the flux tube be $\xi_{\rm CR}\simeq\unit{10}{\%}$ of the supernova ejecta kinetic energy $E_{\text{SNR}}$,
\begin{equation}
    f_0(p)=\frac{\xi_{\rm CR}E_{\rm SNR}}{8\pi^2 a^2 R_{\rm SNR}(t_{\rm rad})\Lambda (m_p{\rm c})^3m_p{\rm c^2}}\left(\frac{p}{m_p{\rm c}}\right)^{-\alpha}.
\end{equation}
Note that the flux tube's radius $a$ is chosen such that the initial volume of the cylindrical CR cloud is the same as that of a sphere with radius $R_{\rm SNR}(t_{\rm rad})$, that is $a=\sqrt{6}R_{\rm SNR}(t_{\rm rad})/3$. $\Lambda$ is a normalisation function,
\begin{eqnarray}
\Lambda=\int^{p_{\rm max}}_{p_{\rm min}}\left(\frac{p}{m_p{\rm c}}\right)^{2-\alpha}\left[\sqrt{\left(\frac{p}{m_p{\rm c}}\right)^2+1}-1\right]\frac{{\rm d}p}{m_p{\rm c}} \, .
\end{eqnarray}
We have adopted $p_{\rm min}=\unit{0.1}{GeV/c}$ ($E_{\rm min}\simeq \unit{5}{MeV}$) and $p_{\rm max}=\unit{5}{PeV/c}$ ($E_{\rm max}\simeq \unit{5}{PeV}$).

The turbulence spectrum is assumed to be the same everywhere initially and follow the background turbulence spectrum defined by a Kraichnan diffusion coefficient 
\begin{equation}
    D_0(p)=\frac{D_B(p)}{kW_{\rm BG}(k)}=\unit{0.03}{\frac{pc^2}{yr}}\lr{\frac{p}{\unit{10}{GeV/c}}}^{1/2}\beta,
\end{equation}
with $k=1/r_{\rm L}(p)$ as in eq.~\eqref{eq:dc}. 
Inside the SNR a higher turbulence level is expected, but as shown by~\cite{Nava_2016}, this has very little impact on the calculations and is therefore omitted here.

\subsection{Boundary conditions}

We consider the flux tube to be symmetric with respect to the position of the SNR. 
It is then sufficient to only consider one half-plane in the coordinate (e.g.\ $z$) along the large-scale field and enforce symmetry. 
Decreasing the advection velocity to zero as $z \to 0$, we can chose a no-flux boundary condition for the phase-space density $f$ at $z=0$, that is $\partial_z f |_{z=0} = 0$. 

The approximation of a 1D flux tube is only valid for scales smaller than the coherence length $L_c$ of the magnetic field. On larger scales, particles diffuse significantly faster in 3D. 
Hence, a free escape boundary is assumed at $z=L_c$.
At high energies, energy losses can be neglected ($t_{\rm loss} \gg t_{\rm diff}$), leading to a power law spectrum.
Therefore, a continuation of the power law spectrum is assumed. 
Since there are no energy gains, a lower boundary condition in momentum is not necessary.

As for the turbulence power spectrum $W$, it retains the background value outside the domain of interest, therefore at both high and small wave numbers continuity of the Kolmogorov/Kraichnan spectrum is assumed, the same at $z=L_c$.

\subsection{Numerical modeling}

To solve the coupled, partial, non-linear differential equations a semi-implicit Crank-Nicolson scheme is used. This allows for significantly bigger time steps compared to explicit schemes oftentimes employed for numerical studies of CR self-confinement. 
The grid is chosen to be finer in regions where a strong gradient in $f$ or $W$ is expected. 
Details can be found in appendix \ref{sec:Numerics}.

\section{Results}
\label{sec:results}

In the following we will show and discuss the results of the non-linear calculation for both the WNM and WIM cases. 
The HIM is not considered for the low-energy particles we are interested in here, as at the late release time $t_{\rm rad}=\unit{200}{kyr}$ and large shock radius $R_{\rm SNR}(t_{\rm rad})=\unit{104}{pc}$ dilute the CR density by a factor of $10$ compared to the WIM case. 
Therefore, the effect of the streaming instability decreases, continuing the trend towards lower energies already observed in~\cite{Nava_2019}.

\subsection{Spatial Distribution}

\begin{figure}[tbp]
\centering
\includegraphics[scale=1]{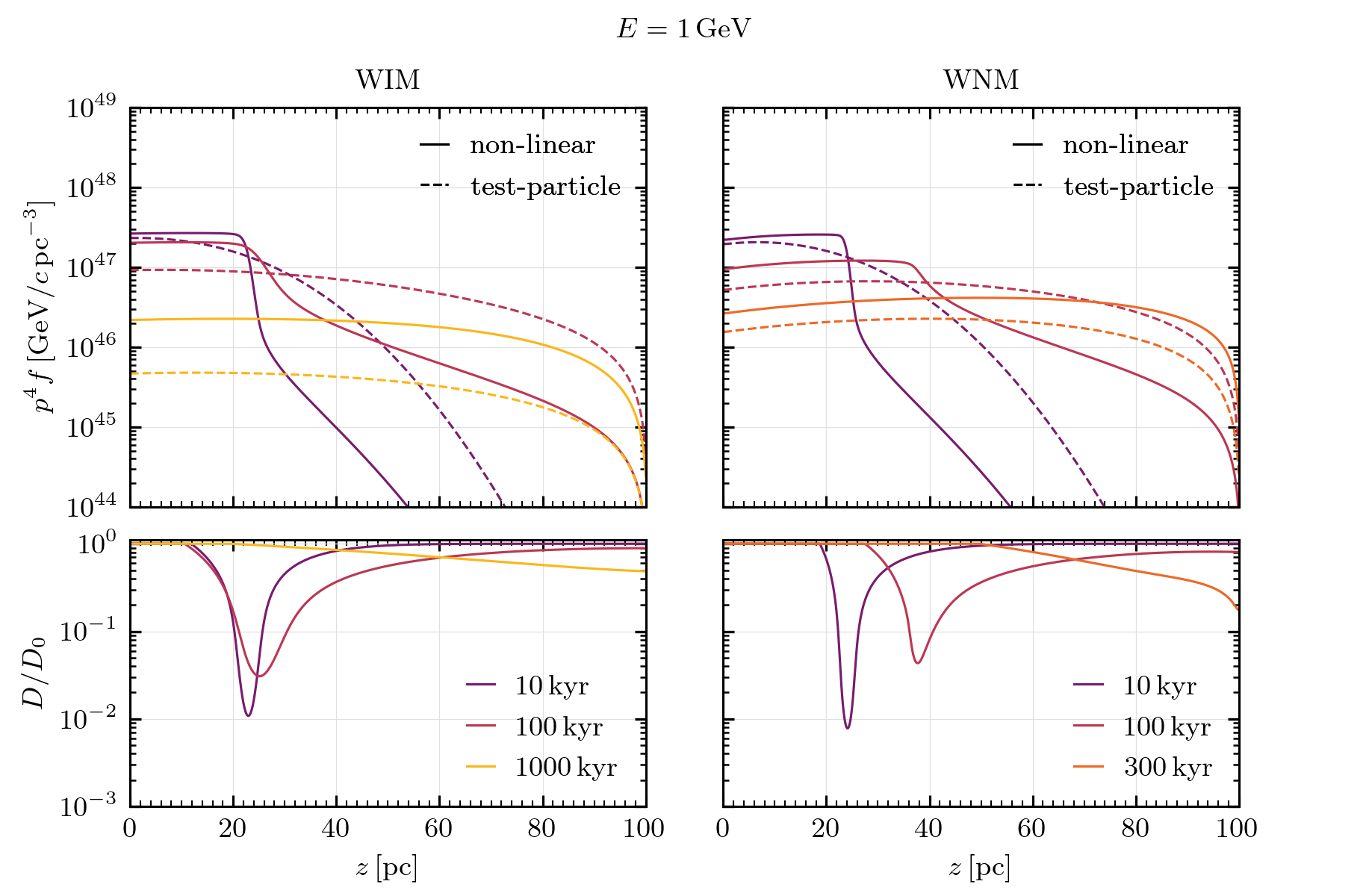}
\caption{\label{fig:z1} Time evolution of the spatial distribution of a CR cloud with $\alpha=4.2$ at $\unit{1}{GeV}$ in the WIM (left) and WNM (right). $z$ is the distance from the supernova. 
Particles are released when the radius of the SNR is $\unit{23}{pc}$, which is reached $\unit{26}{kyr}$ after the supernova explosion. 
All times shown here refer to this release time.
The upper panels show the CR energy density $p^4 f$ with the test-particle solution marked in dashed lines and the non-linear results marked in solid lines.
The lower panels show the corresponding diffusion coefficient $D$ normalised to the background coefficient $D_0$.}
\end{figure}

\begin{figure}[tbp]
\centering
\includegraphics[scale=1]{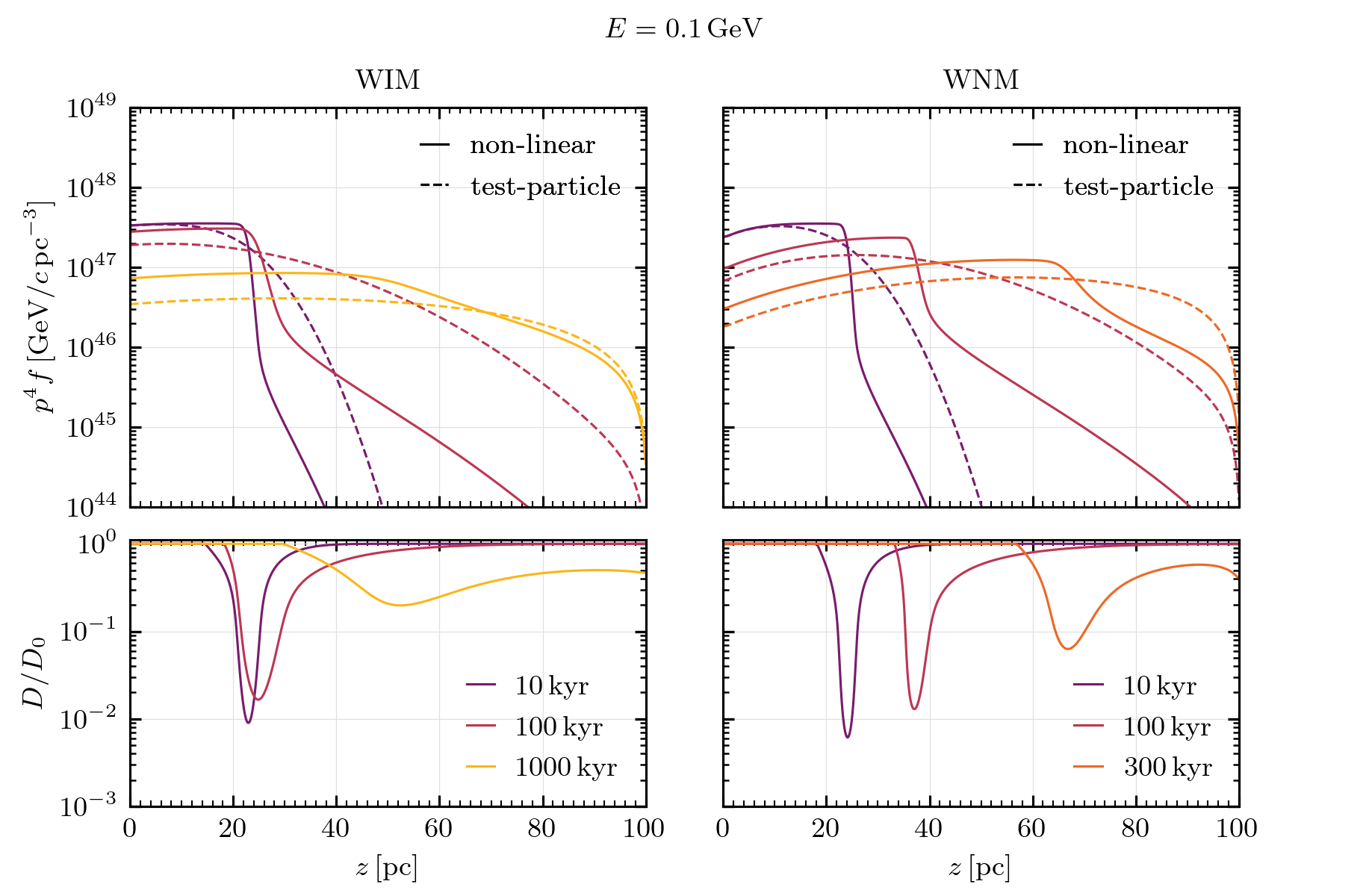}
\caption{\label{fig:z01} Same as fig. \ref{fig:z1}, but for $\unit{0.1}{GeV}$.}
\end{figure}

The spatial distribution of the CR energy density and the suppression of the diffusion coefficient at different times is shown in figure \ref{fig:z1}. 
For the injection spectral index we have assumed $\alpha=4.2$. 
The simulation was started at the begin of the radiative phase, which according to eq. \eqref{eq:trad} happens about $\unit{26}{kyr}$ after the supernova explosion when the SNR has a radius of $\unit{23}{pc}$ (eq. \ref{eq:RST}). 
The upper panels show the CR phase space density $f$ multiplied by $p^4$, that is the differential energy density. 
The dashed lines mark the test-particle solution, which is the solution without backreaction of the turbulent magnetic field (meaning only the background turbulence $W=W_{\rm BG}$ enters the diffusion coefficient). 
The solid lines marks our non-linear model. 
In the lower panel, the diffusion coefficient  is shown with respect to its background value, $D/D_0$.

For the WIM, particles are initially confined behind the SNR shock and we have modelled this with a step-like spatial distribution. 
Once propagating outwards, the spatial distribution of particles resembles the test-particle result, which is an error function in $z$ for times smaller than the growth time of turbulence. 
At these early times the gradient is approximately Gaussian shaped, and therefore the initial decrease of the diffusion coefficient resembles a Gaussian as well. At later times the non-linearity will cause a deviation from this simple shape.
The slower propagation then slows down the expansion of the CR cloud and confines the particles close to the initial SNR. 
A maximum in the suppression of the diffusion coefficient is reached once the turbulent cascade and the damping processes balance the streaming instability.
Energy loss processes and the slow expansion of the particles reduce the streaming instability until the diffusion coefficient returns to the ISM value. 
In contrast to previous results at higher energies \cite{Nava_2016}, the suppression of the diffusion coefficient for particles with energies of around $\unit{100}{MeV}$ lasts more than $\unit{1}{Myr}$ and reaches two orders of magnitude at early times. 
At the outer boundary, particles escape freely into the region where the diffusion is 3D which we have modelled by a free escape boundary condition. 
This creates a stronger gradient again which results in the suppression of the diffusion coefficient at the boundary at later time. 

Concerning the WNM, due to the weak coupling of ions and neutrals at low energies (see section \ref{subsec:damping}) and the low ion number density the Alfvén speed is one order of magnitude above the one obtained for the WIM. 
This will lead not only to a larger advection speed and growth rate rate, but also to a larger transfer in wavenumber and larger damping rates. 
Inside the SNR the phase space density $f$ of particles slightly decreases towards the inner boundary, since particles get advected outwards. 
The large Alfvén speed in the WNM leads to a stronger depletion of particles from the centre of the SNR compared to the WIM case, causing a positive gradient in phase space density inside the SNR. 
Strictly speaking, this would result in backwards propagating waves, but we estimate the dynamical impact of this to be limited and have therefore neglected those. 
We can see that the resulting suppression of the diffusion coefficient is around two orders of magnitude at $\unit{100}{kyr}$ and decreases to one order of magnitude around $\unit{400}{kyr}$. 
At this point, the gradient in particle energy density has propagated to the free escape boundary. Since the velocity at which this happens is faster than the Alfvén speed, especially at later times, particles are not self-confined anymore.
This increases the gradient as particles escape and also leads to a suppression of the diffusion coefficient at the boundary similar to the case of WIM. 
Here, we note again that the sharp transition from 1D to 3D leads to an overestimation of the increase in gradient and therefore, streaming instability compared to a more realistic transition (see also discussions in \cite{Mukhopadhyay_2021}). Since the Alfv\'en speed in the WNM is essentially larger than that in the WIM, the growth rate of turbulence is more strongly overestimated leading to the stronger suppression of the diffusion coefficient at the boundary as compared to the WIM.

\subsection{Spectra}

\begin{figure}[tbp]
\centering
\includegraphics[scale=1]{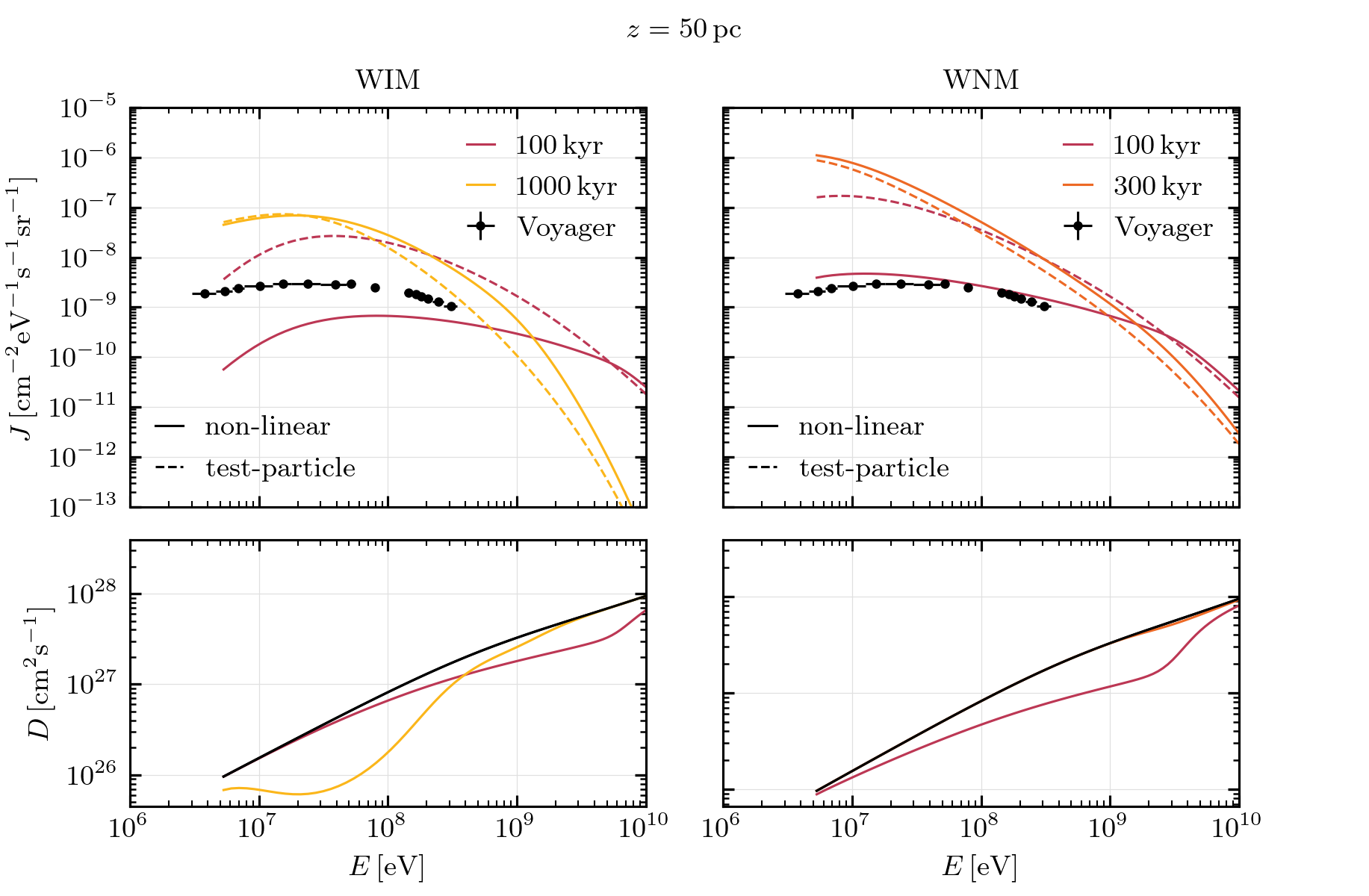}
\caption{\label{fig:E1} CR spectra at a distance of $\unit{50}{pc}$ from the supernova with initial spectral index $\alpha = 4.2$. 
The upper panels show the intensity $J$ and the lower ones the diffusion coefficient $D$, both as a function of kinetic energy $E$. 
Dashed lines mark the test-particle solution and solid ones the non-linear results. 
Both for the WIM (left) and the WNM (right) the spectra are flattened compared to the test-particle result below a few GeV, similar to the observations by Voyager~1 shown as black dots for reference. }
\end{figure}

In figure~\ref{fig:E1}, we show our results as a function of kinetic energy $E$, for a distance of $\unit{50}{pc}$ for both the WIM and WNM. We have assumed the initial spectral index of $\alpha=4.2$. 
The test-particle solution is shown in dashed lines, the result of the non-linear calculation in solid lines. 
In the top panels, we show the proton intensities $J$. 
The black dots mark the intensity measured by Voyager~1~\cite{Voyager_data}. 
In the bottom panels the corresponding diffusion coefficient $D$ is shown. 
The coloured lines show the non-linear result at different times, the black line corresponds to the ISM value. 
Both for the WIM (left panels) and the WNM (right panels) a flattening of the proton spectrum at lower energies is visible. 
Generically, this trend is more favourable for reproducing the Voyager data. 
At higher energies and later times the test-particle solution and the non-linear result agree, since the diffusion coefficient, as shown in the lower panel, has retained the interstellar value again. 
As the diffusion coefficient is still larger at higher energies, the gradient of the phase-space density also propagates out faster for higher energies. 
This leads to earlier self-confinement at higher energies. 
Hence, at some distance from the release radius the diffusion coefficient is first reduced at higher energies, and then gradually at lower ones. 
This leads to a relative reduction of particle intensity at low energies, causing the flattening of the spectrum compared to the test-particle case.
In contrast to the WIM, where the advection speed is small, in the WNM the CR gradient propagates outwards faster, leading to a smaller deviation at later time.

\subsection{Diffusion coefficient}

\begin{figure}[tbp]
\centering
\includegraphics[scale=1]{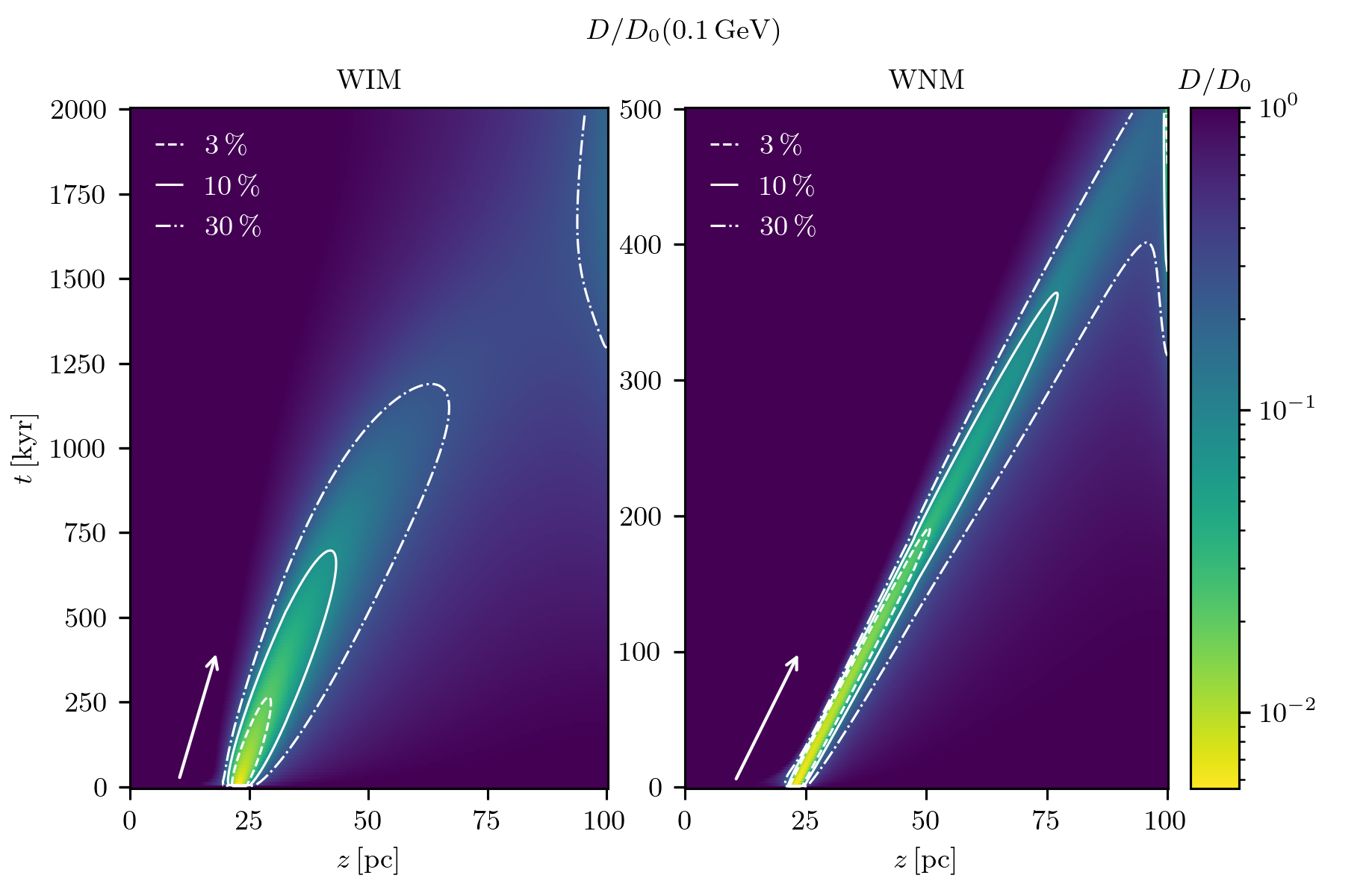}
\caption{\label{fig:DE01} Diffusion coefficient $D$ with respect to its background value $D_0$ as a function of time and space at $\unit{100}{MeV}$. 
The left panel shows the WIM case and the right one the WNM case. 
The contour lines mark different suppression values and the white arrows the local Alfvén speed.
In both cases the suppression is strongest around the edge of the cosmic ray cloud, where the gradient in phase space density is highest. 
Over time both the particles and the turbulence propagates outwards. 
The suppression at the boundary is a result of the transition from 1D to 3D, which is over-estimated here due to the free escape boundary.}
\end{figure}

The difference between the non-linear case and the usual test-particle case is the backreaction on the diffusion coefficient in the former case. 
In order to understand the importance of the suppression of the diffusion coefficient on the flux of CRs it is necessary to have a deeper understanding of its spatial and temporal evolution. 
For an energy of $\unit{100}{MeV}$ this is shown in figure~\ref{fig:DE01}, where the left panel is again for the WIM and the right one for the WNM. 

For both the WIM and the WNM the suppression of the diffusion coefficient is initially strongest at the edge of the SNR at $\unit{23}{pc}$. 
The peak of the suppression is where the gradient is strongest and reaches up to two orders of magnitude. 
With time both the region of the strongest gradient in cosmic rays and the turbulence is advected outwards. 
Due to the weak coupling of neutrals at low energies and a lower ion fraction the Alfvén speed is a factor $5$ larger in the WNM. 
This limits the possible time of suppression to the advection time \mbox{$t_{ \rm adv}=(L_c-R_{\rm SNR})/v_{\rm A}\approx \unit{500}{kyr}$}. 
In the WIM the gradient in CRs is smoothed by the inhibited diffusion faster than the advection time of $\unit{2.5}{Myr}$. 
Therefore, the suppression has subsided at $\unit{1.5}{Myr}$, before it can reach the boundary.
The decreased diffusion at the outer boundary in both cases is a result of the free escape boundary, which has been implemented to approximate the transition form 1D to 3D diffusion. As mentioned above, this transition is unnaturally sharp and therefore leads to an overestimation of the gradient in cosmic rays at the boundary.

\subsection{Grammage}

\begin{figure}[tbp]
\centering
\includegraphics[scale=1]{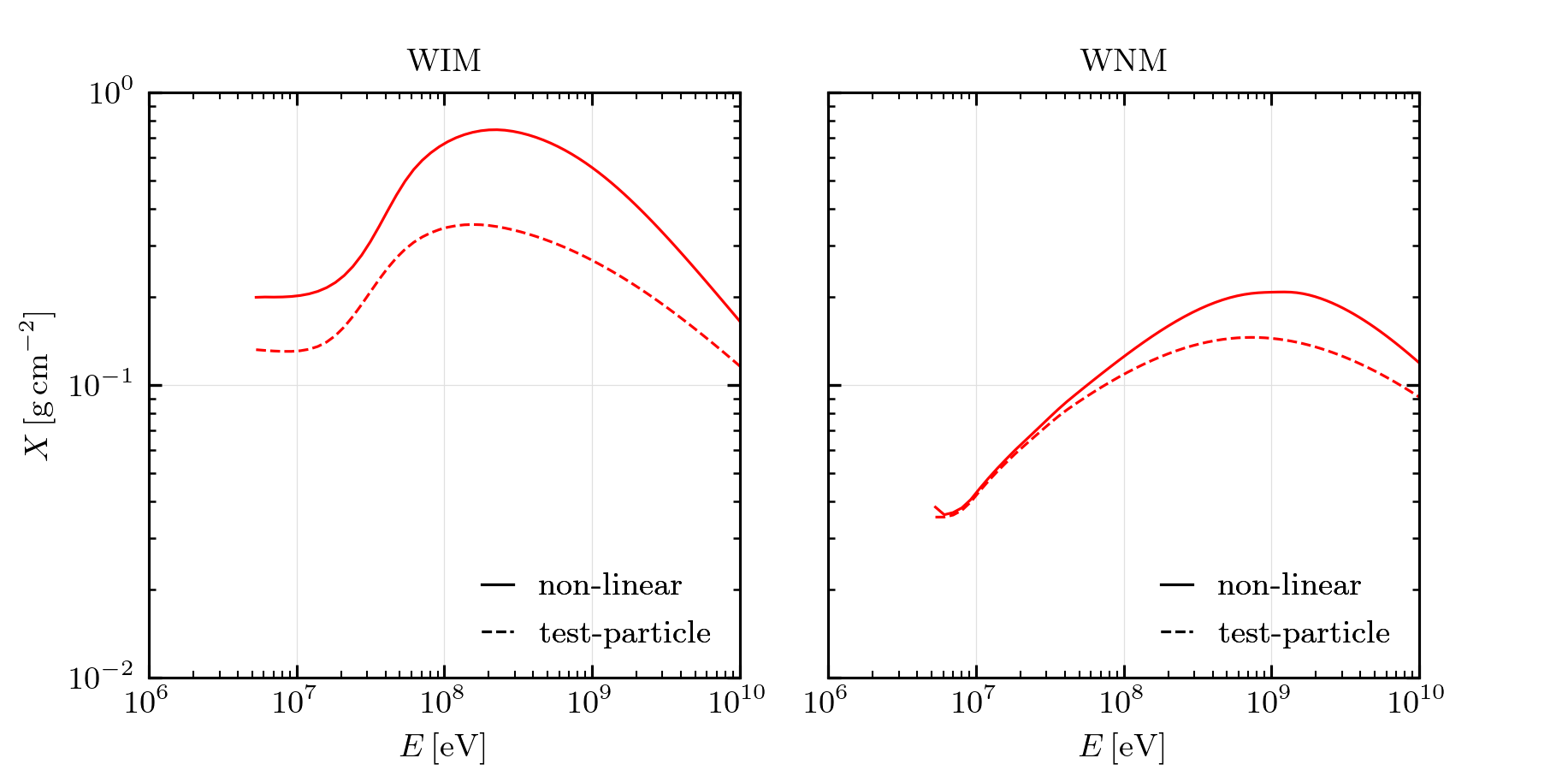}
\caption{\label{fig:Grammage} Grammage $X$ accumulated within the simulation domain as a function of kinetic energy $E$. Left panel: WIM, right panel: WNM. 
The test-particle solution is marked with a dashed line and the result of the non-linear computation with a solid line. 
At $\unit{10}{GeV}$ the grammage agrees with the results of \cite{Recchia_2021}. For WIM the grammage at low energies is increased by up to a factor of 3. 
For the WNM the increase is smaller.}
\end{figure}

During propagation, CRs undergo spallation and produce secondaries upon interacting with matter. The amount of matter traversed is 	parameterised by the grammage which is commonly used in galactic propagation models \cite{grammage_Galactic_Jones}. We shall briefly introduce the approach to estimate the grammage experiences by CRs in the vicinity of the SNR. Let's first identify the grammage of a single particle with momentum $p$ at time $t$ as
\begin{eqnarray}
X_{\rm 1p}(p,t)=\int^t_0\rho v_{0}(p,t'){\rm d} t'.
\end{eqnarray}
where $\rho$ is the mass density, $t$ propagation time $t$, and $v_0=v_0(p,t)$ is the initial speed of particles escaping the simulation region at time $t$ with momentum $p$. Note that we calculate the initial particle's speed from $p_0$ which is the solution of the following equation
\begin{eqnarray}
t=-\int^p_{p_0}\frac{{\rm d} p'}{\left({\rm d}p'/{\rm d}t\right)}.
\end{eqnarray}
We can now define the average source grammage experienced by particles escaping at momentum $p$ as follows
\begin{eqnarray}
\langle X(E)\rangle=\frac{1}{N(p,0)}\int_0^\infty N(p,t)\rho v_0(p,t){\rm d} t,
\end{eqnarray}
where 
\begin{eqnarray}
N(p,t)=\pi a^2 \int^{L_c}_04\pi p^2 f(z,p,t){\rm d}z
\end{eqnarray}
is the total number of CRs with momentum $p$ within the region $0\leq z\leq L_c$ at time $t$. Note again that $a=\sqrt{6}R_{\rm SNR}(t_{\rm rad})/3$ is the radius of the flux tube.

In figure \ref{fig:Grammage} the grammage accumulated in the simulation domain is shown as a function of kinetic energy. 
Dashed lines again mark the test-particle case and the solid lines refer to the non-linear calculation. 
On the left the result in WIM, where the suppression of the diffusion coefficient persists longer and the advection velocity is smaller compared to the WNM, displayed on the right, is shown. 
In the WIM, the inhibited diffusion confines particles close to the SNR for longer times and leads to an increase in the grammage of at most a factor of three compared to the test-particle case. 
At very low energies the smaller velocity and higher energy losses limit the amount of grammage and lead to the decline at $\unit{100}{MeV}$. 
For the WNM, the suppression of the diffusion coefficient only has an effect for high-energy particles, since in the transport at low energies advection dominates and the value of the diffusion coefficient does not matter. 
In  both cases, the grammage accumulated in the near source region is of the order of $\unit{0.1}{g/cm^2}$, which is at most $\unit{10}{\%}$ of the value accumulated in the Galaxy \cite{grammage_Galactic_Gabici,grammage_Galactic_Jones}. At $\unit{10}{GeV}$ the grammage is comparable to \cite{Recchia_2021}.

\section{Conclusion and outlook}
\label{sec:conc}

In this paper we have studied the propagation of low-energy CRs around an SNR.
We considered a cloud of CRs released from the SNR as first proposed by~\cite{Malkov2013}. 
The gradient in CRs gives rise to the resonant streaming instability, creating Alfvénic turbulence upon which they themselves scatter. 
This makes the problem highly non-linear and difficult to solve analytically. 
Previous studies have considered particles with energies above $\unit{10}{GeV}$, however most of the energy density in CRs is stored in particles around $\unit{1}{GeV}$. 
Additionally, with the Voyager probes we now have direct measurements of the intensities of low-energy CRs unhindered by solar modulation. 
Therefore, we expanded previous studies by \cite{Nava_2016,Nava_2019,Brahimi2020,Recchia_2021} to lower energies, where energy losses become important.

The main result of this study is that the diffusion coefficient around $\unit{100}{MeV}$ can be suppressed for more than $\unit{1}{Myr}$ in the WIM and due to the higher advection speed up to $\unit{300}{kyr}$ in the WNM. 
This is significantly longer than found at higher energies. In both cases the inhibition initially reaches two orders of magnitude. 

We find also that the non-linear spectrum flattens below $\unit{1}{GeV}$ similar to the observations of Voyager \cite{Voyager_data}, in contrast to the test-particle solution. 
This mechanism might help to better explain the Voyager intensity in models for CRs from stochastic sources as investigated in \cite{Phan_2021}. 
The self-confinement of CRs might also be responsible for the surprisingly high ionisation rates in molecular clouds in the vicinity of several SNRs, e.g. IC443, W51, or W28 \cite{Indriolo_2010,Ceccarelli_2011,Vaupre_2014,Phan_2020}.

Interestingly, the grammage accumulated in the simulation domain is at most $\unit{0.8}{g/cm^{2}}$, which is an increase by a factor of $3$ compared to the test-particle case. 
This is small compared to the inferred grammage obtained from the ratio of secondary-primary Galactic CRs, which is of the order of $\unit{10}{g/cm^{2}}$ at $\unit{1}{GeV}$~\cite{grammage_Galactic_Gabici}. 
It might be, however, important for precision fitting of the CR spectra \cite{grammage,Genolini_2021}.  

A self consistent treatment of non-linear CR acceleration, escape and propagation is computationally not feasible at the current time and will remain for future work.
The validity of the model used here depends on the value of the magnetic coherence length $L_c$, for smaller values the 1D approximation will break down at smaller distances form the supernova. 
It can already be seen in figure \ref{fig:z01} that the sharp transition from 1D to 3D assumed here is an oversimplification.
A more comprehensive model of the transport in the near source region must consider this transition. Our model however can be used as a very useful limiting case and our results remain robust as long as the coherence length is not altered significantly.

\appendix
\section{Numerical technique}
\label{sec:Numerics}

An analytic solution of the coupled, partial, non-linear differential equations describing the transport of cosmic rays and turbulent power is not possible. Therefore, they are solved numerically on a non-uniform finite grid using second order accurate central differences.
Casting the differential equation in the form $\partial f / \partial t = \mathcal{L} f$, this can be done either by evaluating $\mathcal{L} f$ at $t$ when computing ($t + \Delta t$), called explicit, or at ($t + \Delta t$), called implicit, or by averaging both, called semi-implicit.
Here a semi-implicit Crank-Nicolson (CN) \cite{Crank_1947} scheme is used, which has the advantage that the Courant–Friedrichs–Lewy stability condition \cite{Courant_1928}, which sets an upper limit on $\Delta t$ does not apply. 
Instead, for a simple advection-diffusion problem the scheme is unconditionally stable \cite{numericalrecipies}. 
In a one dimensional case the CN scheme results in a tridiagonal matrix equation, which can efficiently be solved using the Thomas algorithm \cite{thomas}. 
Unlike previous work, the equations at different momenta are not independent, but coupled by energy losses and the turbulent cascade. 
The energy, which was a parameter before is now an additional independent variable. 
Even though the CN scheme is still applicable the matrix will be sparse but not tridiagonal. 
To prevent this the individual equations are split into a part depending only on differential operators in space and another one in energy. This is called operator splitting and more comprehensively explained in Chapter 20.3.2 of Ref.~\cite{numericalrecipies}. 
The coupling between the phase space density and the turbulent spectral power is done in an explicit way. 
Also the non-linear Landau damping is dealt with in an explicit way.
Both of the above can make the code unstable, but we find that the time steps are still significantly larger than allowed in a purely explicit scheme.
In order to linearise the turbulent cascade, the method explained in \cite{numericalrecipies} chapter 20.2 (p.1047f) is used.

The grids are chosen to be finer in regions, where a strong gradient in $f$ or $W$ is expected. 
The momentum grid extends from $p_{\text{min}}=\unit{100}{MeV}$ to $p_{\text{max}}=\unit{10}{GeV}$ and is logarithmic with $33$ bins per decade.
The grid for the power spectrum $W$ is resonant to this grid, meaning there is a one-to-one correspondence $k = 1/r_{\text{L}}$.
The spatial grid depends on the phase of the ISM, since $v_{A,i}$, relevant at low energies, depends on $n_i$.
In total $1000$ spatial grid points are chosen, $2/3$ of which are equidistant between $\unit{R_{\text{SN}}(t_{\text{rad}})-10}{pc}$ and  $\unit{R_{\text{SN}}(t_{\text{rad}})+10}{pc}$ for the WIM and $\unit{R_{\text{SN}}(t_{\text{rad}})-10}{pc}$ and $\unit{R_{\text{SN}}(t_{\text{rad}})+25}{pc}$ for the WNM. The other $1/3$ of the grid points are equidistant on the rest of the domain. In all cases the timestep is $\Delta t=\unit{5}{yr}$.

\acknowledgments
The authors would like to thank Sarah Recchia, Carmelo Evoli, Tim Linden, Giovanni Morlino and Stefano Gabici for fruitful discussions. HJ also thanks Marco Kuhlen for many interesting conversations.

\bibliographystyle{JHEP}
\bibliography{main}

\providecommand{\href}[2]{#2}\begingroup\raggedright\begin{thebibliography}{10}

\bibitem{ZwickyBaade}
W.~{Baade} and F.~{Zwicky}, \emph{{Cosmic Rays from Super-novae}},
  \href{https://doi.org/10.1073/pnas.20.5.259}{\emph{Proceedings of the
  National Academy of Science} {\bfseries 20} (1934) 259}.

\bibitem{Fermi-LAT:2013iui}
{\scshape Fermi-LAT} collaboration, \emph{{Detection of the Characteristic
  Pion-Decay Signature in Supernova Remnants}},
  \href{https://doi.org/10.1126/science.1231160}{\emph{Science} {\bfseries 339}
  (2013) 807} [\href{https://arxiv.org/abs/1302.3307}{{\ttfamily 1302.3307}}].

\bibitem{Jogler:2016lav}
T.~Jogler and S.~Funk, \emph{{Revealing W51c as a Cosmic ray Source Using
  Fermi-lat Data}},
  \href{https://doi.org/10.3847/0004-637X/816/2/100}{\emph{Astrophys. J.}
  {\bfseries 816} (2016) 100}.

\bibitem{grammage_Galactic_Gabici}
S.~Gabici, C.~Evoli, D.~Gaggero, P.~Lipari, P.~Mertsch, E.~Orlando et~al.,
  \emph{The origin of galactic cosmic rays: Challenges to the standard
  paradigm},
  \href{https://doi.org/10.1142/s0218271819300222}{\emph{International Journal
  of Modern Physics D} {\bfseries 28} (2019) 1930022}.

\bibitem{Kulsrud+2020}
R.M.~Kulsrud, \emph{Plasma Physics for Astrophysics}, Princeton University
  Press (2020),
  \href{https://doi.org/doi:10.1515/9780691213354}{doi:10.1515/9780691213354}.

\bibitem{Marcowith_2021}
A.~{Marcowith}, A.J.~{van Marle} and I.~{Plotnikov}, \emph{{The cosmic
  ray-driven streaming instability in astrophysical and space plasmas}},
  \href{https://doi.org/10.1063/5.0013662}{\emph{Physics of Plasmas} {\bfseries
  28} (2021) 080601}.

\bibitem{Bell_2004}
A.R.~{Bell}, \emph{{Turbulent amplification of magnetic field and diffusive
  shock acceleration of cosmic rays}},
  \href{https://doi.org/10.1111/j.1365-2966.2004.08097.x}{\emph{\mnras}
  {\bfseries 353} (2004) 550}.

\bibitem{Amato_2009}
E.~{Amato} and P.~{Blasi}, \emph{{A kinetic approach to cosmic-ray-induced
  streaming instability at supernova shocks}},
  \href{https://doi.org/10.1111/j.1365-2966.2008.14200.x}{\emph{\mnras}
  {\bfseries 392} (2009) 1591}
  [\href{https://arxiv.org/abs/0806.1223}{{\ttfamily 0806.1223}}].

\bibitem{Malkov2013}
M.A.~Malkov, P.H.~Diamond, R.Z.~Sagdeev, F.A.~Aharonian and I.V.~Moskalenko,
  \emph{{Analytic solution for self-regulated collective escape of cosmic rays
  from their acceleration sites}},
  \href{https://doi.org/10.1088/0004-637X/768/1/73}{\emph{Astrophysical
  Journal} {\bfseries 768} (2013) 1}
  [\href{https://arxiv.org/abs/1207.4728}{{\ttfamily 1207.4728}}].

\bibitem{Nava_2016}
L.~Nava, S.~Gabici, A.~Marcowith, G.~Morlino and V.S.~Ptuskin, \emph{Non-linear
  diffusion of cosmic rays escaping from supernova remnants – i. the effect
  of neutrals}, \href{https://doi.org/10.1093/mnras/stw1592}{\emph{Monthly
  Notices of the Royal Astronomical Society} {\bfseries 461} (2016)
  3552–3562}.

\bibitem{Nava_2019}
L.~Nava, S.~Recchia, S.~Gabici, A.~Marcowith, L.~Brahimi and V.~Ptuskin,
  \emph{Non-linear diffusion of cosmic rays escaping from supernova remnants
  – ii. hot ionized media},
  \href{https://doi.org/10.1093/mnras/stz137}{\emph{Monthly Notices of the
  Royal Astronomical Society} {\bfseries 484} (2019) 2684–2691}.

\bibitem{Brahimi2020}
L.~Brahimi, A.~Marcowith and V.S.~Ptuskin, \emph{{Nonlinear diffusion of cosmic
  rays escaping from supernova remnants: Cold partially neutral atomic and
  molecular phases}},
  \href{https://doi.org/10.1051/0004-6361/201936166}{\emph{Astronomy and
  Astrophysics} {\bfseries 633} (2020) }
  [\href{https://arxiv.org/abs/1909.04530}{{\ttfamily 1909.04530}}].

\bibitem{Recchia_2021}
S.~{Recchia}, D.~{Galli}, L.~{Nava}, M.~{Padovani}, S.~{Gabici}, A.~{Marcowith}
  et~al., \emph{{Grammage of cosmic rays in the proximity of supernova remnants
  embedded in a partially ionized medium}}, {\emph{arXiv e-prints} (2021)
  arXiv:2106.04948} [\href{https://arxiv.org/abs/2106.04948}{{\ttfamily
  2106.04948}}].

\bibitem{Voyager_out}
S.M.~{Krimigis}, R.B.~{Decker}, E.C.~{Roelof}, M.E.~{Hill}, T.P.~{Armstrong},
  G.~{Gloeckler} et~al., \emph{{Search for the Exit: Voyager 1 at
  Heliosphere{\textquoteright}s Border with the Galaxy}},
  \href{https://doi.org/10.1126/science.1235721}{\emph{Science} {\bfseries 341}
  (2013) 144}.

\bibitem{Voyager_data}
A.C.~{Cummings}, E.C.~{Stone}, B.C.~{Heikkila}, N.~{Lal}, W.R.~{Webber},
  G.~{J{\'o}hannesson} et~al., \emph{{Galactic Cosmic Rays in the Local
  Interstellar Medium: Voyager 1 Observations and Model Results}},
  \href{https://doi.org/10.3847/0004-637X/831/1/18}{\emph{\apj} {\bfseries 831}
  (2016) 18}.

\bibitem{Boschini_2018a}
M.J.~{Boschini}, S.~{Della Torre}, M.~{Gervasi}, D.~{Grandi},
  G.~{J{\'o}hannesson}, G.~{La Vacca} et~al., \emph{{HelMod in the Works: From
  Direct Observations to the Local Interstellar Spectrum of Cosmic-Ray
  Electrons}}, \href{https://doi.org/10.3847/1538-4357/aaa75e}{\emph{\apj}
  {\bfseries 854} (2018) 94}
  [\href{https://arxiv.org/abs/1801.04059}{{\ttfamily 1801.04059}}].

\bibitem{Boschini_2018b}
M.J.~{Boschini}, S.~{Della Torre}, M.~{Gervasi}, D.~{Grandi},
  G.~{J{\'o}hannesson}, G.~{La Vacca} et~al., \emph{{Deciphering the Local
  Interstellar Spectra of Primary Cosmic-Ray Species with HELMOD}},
  \href{https://doi.org/10.3847/1538-4357/aabc54}{\emph{\apj} {\bfseries 858}
  (2018) 61} [\href{https://arxiv.org/abs/1804.06956}{{\ttfamily 1804.06956}}].

\bibitem{Orlando_2018}
E.~{Orlando}, \emph{{Imprints of cosmic rays in multifrequency observations of
  the interstellar emission}},
  \href{https://doi.org/10.1093/mnras/stx3280}{\emph{\mnras} {\bfseries 475}
  (2018) 2724} [\href{https://arxiv.org/abs/1712.07127}{{\ttfamily
  1712.07127}}].

\bibitem{source_break}
A.~Vittino, P.~Mertsch, H.~Gast and S.~Schael, \emph{Breaks in interstellar
  spectra of positrons and electrons derived from time-dependent ams data},
  \href{https://doi.org/10.1103/physrevd.100.043007}{\emph{Physical Review D}
  {\bfseries 100} (2019) }.

\bibitem{Bisschoff_2019}
D.~{Bisschoff}, M.S.~{Potgieter} and O.P.M.~{Aslam}, \emph{{New Very Local
  Interstellar Spectra for Electrons, Positrons, Protons, and Light Cosmic Ray
  Nuclei}}, \href{https://doi.org/10.3847/1538-4357/ab1e4a}{\emph{\apj}
  {\bfseries 878} (2019) 59}
  [\href{https://arxiv.org/abs/1902.10438}{{\ttfamily 1902.10438}}].

\bibitem{Phan_2021}
V.H.M.~Phan, F.~Schulze, P.~Mertsch, S.~Recchia and S.~Gabici, \emph{Stochastic
  fluctuations of low-energy cosmic rays and the interpretation of voyager
  data}, \href{https://doi.org/10.1103/PhysRevLett.127.141101}{\emph{Phys. Rev.
  Lett.} {\bfseries 127} (2021) 141101}.

\bibitem{Padovani_2009}
M.~{Padovani}, D.~{Galli} and A.E.~{Glassgold}, \emph{{Cosmic-ray ionization of
  molecular clouds}},
  \href{https://doi.org/10.1051/0004-6361/200911794}{\emph{\aap} {\bfseries
  501} (2009) 619} [\href{https://arxiv.org/abs/0904.4149}{{\ttfamily
  0904.4149}}].

\bibitem{Morlino_2015}
G.~{Morlino} and S.~{Gabici}, \emph{{Cosmic ray penetration in diffuse
  clouds.}}, \href{https://doi.org/10.1093/mnrasl/slv074}{\emph{\mnras}
  {\bfseries 451} (2015) L100}
  [\href{https://arxiv.org/abs/1503.02435}{{\ttfamily 1503.02435}}].

\bibitem{Ivlev_2015}
A.V.~{Ivlev}, M.~{Padovani}, D.~{Galli} and P.~{Caselli}, \emph{{Interstellar
  Dust Charging in Dense Molecular Clouds: Cosmic Ray Effects}},
  \href{https://doi.org/10.1088/0004-637X/812/2/135}{\emph{\apj} {\bfseries
  812} (2015) 135} [\href{https://arxiv.org/abs/1507.00692}{{\ttfamily
  1507.00692}}].

\bibitem{Phan_2018}
V.H.M.~{Phan}, G.~{Morlino} and S.~{Gabici}, \emph{{What causes the ionization
  rates observed in diffuse molecular clouds? The role of cosmic ray protons
  and electrons}}, \href{https://doi.org/10.1093/mnras/sty2235}{\emph{\mnras}
  {\bfseries 480} (2018) 5167}
  [\href{https://arxiv.org/abs/1804.10106}{{\ttfamily 1804.10106}}].

\bibitem{Indriolo_2010}
N.~{Indriolo}, G.A.~{Blake}, M.~{Goto}, T.~{Usuda}, T.~{Oka}, T.R.~{Geballe}
  et~al., \emph{{Investigating the Cosmic-ray Ionization Rate Near the
  Supernova Remnant IC 443 through H$^{+}$ $_{3}$ Observations}},
  \href{https://doi.org/10.1088/0004-637X/724/2/1357}{\emph{\apj} {\bfseries
  724} (2010) 1357} [\href{https://arxiv.org/abs/1010.3252}{{\ttfamily
  1010.3252}}].

\bibitem{Ceccarelli_2011}
C.~{Ceccarelli}, P.~{Hily-Blant}, T.~{Montmerle}, G.~{Dubus}, Y.~{Gallant} and
  A.~{Fiasson}, \emph{{Supernova-enhanced Cosmic-Ray Ionization and Induced
  Chemistry in a Molecular Cloud of W51C}},
  \href{https://doi.org/10.1088/2041-8205/740/1/L4}{\emph{\apjl} {\bfseries
  740} (2011) L4} [\href{https://arxiv.org/abs/1108.3600}{{\ttfamily
  1108.3600}}].

\bibitem{Vaupre_2014}
S.~Vaupr{\'{e}}, P.~Hily-Blant, C.~Ceccarelli, G.~Dubus, S.~Gabici and
  T.~Montmerle, \emph{{Cosmic ray induced ionisation of a molecular cloud
  shocked by the W28 supernova remnant}},
  \href{https://doi.org/10.1051/0004-6361/201424036}{\emph{Astronomy and
  Astrophysics} {\bfseries 568} (2014) }
  [\href{https://arxiv.org/abs/1407.0205}{{\ttfamily 1407.0205}}].

\bibitem{Phan_2020}
V.H.M.~{Phan}, S.~{Gabici}, G.~{Morlino}, R.~{Terrier}, J.~{Vink}, J.~{Krause}
  et~al., \emph{{Constraining the cosmic ray spectrum in the vicinity of the
  supernova remnant W28: from sub-GeV to multi-TeV energies}},
  \href{https://doi.org/10.1051/0004-6361/201936927}{\emph{\aap} {\bfseries
  635} (2020) A40} [\href{https://arxiv.org/abs/1910.09987}{{\ttfamily
  1910.09987}}].

\bibitem{LC_Beck}
M.C.~{Beck}, A.M.~{Beck}, R.~{Beck}, K.~{Dolag}, A.W.~{Strong} and
  P.~{Nielaba}, \emph{{New constraints on modelling the random magnetic field
  of the MW}},
  \href{https://doi.org/10.1088/1475-7516/2016/05/056}{\emph{\jcap} {\bfseries
  2016} (2016) 056} [\href{https://arxiv.org/abs/1409.5120}{{\ttfamily
  1409.5120}}].

\bibitem{Jokipii_1966}
J.R.~{Jokipii}, \emph{{Cosmic-Ray Propagation. I. Charged Particles in a Random
  Magnetic Field}}, \href{https://doi.org/10.1086/148912}{\emph{\apj}
  {\bfseries 146} (1966) 480}.

\bibitem{Blandford_1987}
R.~{Blandford} and D.~{Eichler}, \emph{{Particle acceleration at astrophysical
  shocks: A theory of cosmic ray origin}},
  \href{https://doi.org/10.1016/0370-1573(87)90134-7}{\emph{\physrep}
  {\bfseries 154} (1987) 1}.

\bibitem{Bell_1978}
A.R.~{Bell}, \emph{{The acceleration of cosmic rays in shock fronts - I.}},
  \href{https://doi.org/10.1093/mnras/182.2.147}{\emph{\mnras} {\bfseries 182}
  (1978) 147}.

\bibitem{TEW}
J.A.~{Eilek}, \emph{{Particle reacceleration in radio galaxies.}},
  \href{https://doi.org/10.1086/157093}{\emph{\apj} {\bfseries 230} (1979)
  373}.

\bibitem{Evoli_2018}
C.~{Evoli}, T.~{Linden} and G.~{Morlino}, \emph{{Self-generated cosmic-ray
  confinement in TeV halos: Implications for TeV {\ensuremath{\gamma}} -ray
  emission and the positron excess}},
  \href{https://doi.org/10.1103/PhysRevD.98.063017}{\emph{\prd} {\bfseries 98}
  (2018) 063017} [\href{https://arxiv.org/abs/1807.09263}{{\ttfamily
  1807.09263}}].

\bibitem{Mukhopadhyay_2021}
P.~{Mukhopadhyay} and T.~{Linden}, \emph{{Self-Generated Cosmic-Ray Turbulence
  Can Explain the Morphology of TeV Halos}}, {\emph{arXiv e-prints} (2021)
  arXiv:2111.01143} [\href{https://arxiv.org/abs/2111.01143}{{\ttfamily
  2111.01143}}].

\bibitem{DCk_Miller}
J.A.~{Miller} and D.A.~{Roberts}, \emph{{Stochastic Proton Acceleration by
  Cascading Alfven Waves in Impulsive Solar Flares}},
  \href{https://doi.org/10.1086/176359}{\emph{\apj} {\bfseries 452} (1995)
  912}.

\bibitem{Eloss_c_i}
K.~{Mannheim} and R.~{Schlickeiser}, \emph{{Interactions of cosmic ray
  nuclei}}, {\emph{\aap} {\bfseries 286} (1994) 983}
  [\href{https://arxiv.org/abs/astro-ph/9402042}{{\ttfamily
  astro-ph/9402042}}].

\bibitem{Eloss_pp}
S.~{Krakau} and R.~{Schlickeiser}, \emph{{Pion Production Momentum Loss of
  Cosmic Ray Hadrons}},
  \href{https://doi.org/10.1088/0004-637X/802/2/114}{\emph{\apj} {\bfseries
  802} (2015) 114}.

\bibitem{Krall_Trivelpiece}
N.A.~Krall and A.W.~Trivelpiece, \emph{Principles of plasma physics},
  \href{https://doi.org/10.1119/1.1987587}{\emph{American Journal of Physics}
  {\bfseries 41} (1973) 1380}
  [\href{https://arxiv.org/abs/https://doi.org/10.1119/1.1987587}{{\ttfamily
  https://doi.org/10.1119/1.1987587}}].

\bibitem{ressi_protons_Achterberg}
A.~{Achterberg}, \emph{{Modification of scattering waves and its importance for
  shock acceleration}}, {\emph{\aap} {\bfseries 119} (1983) 274}.

\bibitem{1971ApJ...170..265S}
J.~{Skilling}, \emph{{Cosmic Rays in the Galaxy: Convection or Diffusion?}},
  \href{https://doi.org/10.1086/151210}{\emph{\apj} {\bfseries 170} (1971)
  265}.

\bibitem{ISM_2001}
K.M.~{Ferri{\`e}re}, \emph{{The interstellar environment of our galaxy}},
  \href{https://doi.org/10.1103/RevModPhys.73.1031}{\emph{Reviews of Modern
  Physics} {\bfseries 73} (2001) 1031}
  [\href{https://arxiv.org/abs/astro-ph/0106359}{{\ttfamily
  astro-ph/0106359}}].

\bibitem{ISM_2019}
K.~{Ferri{\`e}re}, \emph{{Plasma turbulence in the interstellar medium}},
  \href{https://doi.org/10.1088/1361-6587/ab49eb}{\emph{Plasma Physics and
  Controlled Fusion} {\bfseries 62} (2020) 014014}
  [\href{https://arxiv.org/abs/1912.08237}{{\ttfamily 1912.08237}}].

\bibitem{ISM_CR_COMP}
V.~{Tatischeff}, J.C.~{Raymond}, J.~{Duprat}, S.~{Gabici} and S.~{Recchia},
  \emph{{The origin of Galactic cosmic rays as revealed by their composition}},
  \href{https://doi.org/10.1093/mnras/stab2533}{\emph{\mnras} {\bfseries 508}
  (2021) 1321} [\href{https://arxiv.org/abs/2106.15581}{{\ttfamily
  2106.15581}}].

\bibitem{BGturb}
M.-M.~Mac~Low and R.S.~Klessen, \emph{Control of star formation by supersonic
  turbulence}, \href{https://doi.org/10.1103/RevModPhys.76.125}{\emph{Rev. Mod.
  Phys.} {\bfseries 76} (2004) 125}.

\bibitem{FarmerGoldreich}
A.J.~Farmer and P.~Goldreich, \emph{Wave damping by magnetohydrodynamic
  turbulence and its effect on cosmic‐ray propagation in the interstellar
  medium}, \href{https://doi.org/10.1086/382040}{\emph{The Astrophysical
  Journal} {\bfseries 604} (2004) 671–674}.

\bibitem{YanLazarian}
H.~{Yan} and A.~{Lazarian}, \emph{{Cosmic-Ray Scattering and Streaming in
  Compressible Magnetohydrodynamic Turbulence}},
  \href{https://doi.org/10.1086/423733}{\emph{\apj} {\bfseries 614} (2004) 757}
  [\href{https://arxiv.org/abs/astro-ph/0408172}{{\ttfamily
  astro-ph/0408172}}].

\bibitem{nlld}
J.F.~{McKenzie} and R.A.B.~{Bond}, \emph{{The role of non-linear Landau damping
  in cosmic ray shock acceleration}}, {\emph{\aap} {\bfseries 123} (1983) 111}.

\bibitem{SNRfree}
R.A.~{Chevalier}, \emph{{Self-similar solutions for the interaction of stellar
  ejecta with an external medium.}},
  \href{https://doi.org/10.1086/160126}{\emph{\apj} {\bfseries 258} (1982)
  790}.

\bibitem{R_ST}
J.K.~Truelove and C.F.~McKee, \emph{Evolution of nonradiative supernova
  remnants}, \href{https://doi.org/10.1086/313176}{\emph{The Astrophysical
  Journal Supplement Series} {\bfseries 120} (1999) 299}.

\bibitem{t_rad}
D.F.~{Cioffi}, C.F.~{McKee} and E.~{Bertschinger}, \emph{{Dynamics of Radiative
  Supernova Remnants}}, \href{https://doi.org/10.1086/166834}{\emph{\apj}
  {\bfseries 334} (1988) 252}.

\bibitem{Malkov_Drury_2001}
M.A.~{Malkov} and L.O.~{Drury}, \emph{{Nonlinear theory of diffusive
  acceleration of particles by shock waves}},
  \href{https://doi.org/10.1088/0034-4885/64/4/201}{\emph{Reports on Progress
  in Physics} {\bfseries 64} (2001) 429}.

\bibitem{nldsa_diesing_2021}
R.~{Diesing} and D.~{Caprioli}, \emph{{Steep Cosmic-Ray Spectra with Revised
  Diffusive Shock Acceleration}},
  \href{https://doi.org/10.3847/1538-4357/ac22fe}{\emph{\apj} {\bfseries 922}
  (2021) 1} [\href{https://arxiv.org/abs/2107.08520}{{\ttfamily 2107.08520}}].

\bibitem{grammage_Galactic_Jones}
F.C.~Jones, A.~Lukasiak, V.~Ptuskin and W.~Webber, \emph{The modified weighted
  slab technique: Models and results},
  \href{https://doi.org/10.1086/318358}{\emph{The Astrophysical Journal}
  {\bfseries 547} (2001) 264}.

\bibitem{grammage}
C.~Evoli, R.~Aloisio and P.~Blasi, \emph{Galactic cosmic rays after the ams-02
  observations},
  \href{https://doi.org/10.1103/physrevd.99.103023}{\emph{Physical Review D}
  {\bfseries 99} (2019) }.

\bibitem{Genolini_2021}
Y.~{G{\'e}nolini}, M.~{Boudaud}, M.~{Cirelli}, L.~{Derome}, J.~{Lavalle},
  D.~{Maurin} et~al., \emph{{New minimal, median, and maximal propagation
  models for dark matter searches with Galactic cosmic rays}},
  \href{https://doi.org/10.1103/PhysRevD.104.083005}{\emph{\prd} {\bfseries
  104} (2021) 083005} [\href{https://arxiv.org/abs/2103.04108}{{\ttfamily
  2103.04108}}].

\bibitem{Crank_1947}
J.~{Crank}, P.~{Nicolson} and D.R.~{Hartree}, \emph{{A practical method for
  numerical evaluation of solutions of partial differential equations of the
  heat-conduction type}},
  \href{https://doi.org/10.1017/S0305004100023197}{\emph{Proceedings of the
  Cambridge Philosophical Society} {\bfseries 43} (1947) 50}.

\bibitem{Courant_1928}
R.~{Courant}, K.~{Friedrichs} and H.~{Lewy}, \emph{{{\"U}ber die partiellen
  Differenzengleichungen der mathematischen Physik}},
  \href{https://doi.org/10.1007/BF01448839}{\emph{Mathematische Annalen}
  {\bfseries 100} (1928) 32}.

\bibitem{numericalrecipies}
W.~Press, S.~Teukolsky, W.~Vetterling and B.~Flannery, \emph{Numerical Recipes:
  The Art of Scientific Computing}, Cambridge University Press, 3~ed. (2007).

\bibitem{thomas}
L.~Thomas, \emph{{Elliptic Problems in Linear Difference Equations over a
  Network}}, {\emph{Watson Sc. Comp. Lab. Rep., Columbia University, New York.}
  (1949) }.

\end{thebibliography}\endgroup

\end{document}